\input harvmac
\noblackbox
\newcount\figno
\figno=0
\def\fig#1#2#3{
\par\begingroup\parindent=0pt\leftskip=1cm\rightskip=1cm\parindent=0pt
\baselineskip=11pt
\global\advance\figno by 1
\midinsert
\epsfxsize=#3
\centerline{\epsfbox{#2}}
\vskip 12pt
\centerline{{\bf Figure \the\figno:} #1}\par
\endinsert\endgroup\par}
\def\figlabel#1{\xdef#1{\the\figno}}

\def\np#1#2#3{Nucl. Phys. {\bf B#1} (#2) #3}
\def\pl#1#2#3{Phys. Lett. {\bf B#1} (#2) #3}
\def\prl#1#2#3{Phys. Rev. Lett. {\bf #1} (#2) #3}
\def\prd#1#2#3{Phys. Rev. {\bf D#1} (#2) #3}

\def\cmp#1#2#3{Comm. Math. Phys. {\bf #1} (#2) #3}


\font\cmss=cmss10
\font\cmsss=cmss10 at 7pt
\def\rlx{\relax\leavevmode}
\def\inbar{\vrule height1.5ex width.4pt depth0pt}
\def\IC{\relax\,\hbox{$\inbar\kern-.3em{\rm C}$}}
\def\IN{\relax{\rm I\kern-.18em N}}
\def\IP{\relax{\rm I\kern-.18em P}}
\def\ZZ{\rlx\leavevmode\ifmmode\mathchoice{\hbox{\cmss Z\kern-.4em Z}}
 {\hbox{\cmss Z\kern-.4em Z}}{\lower.9pt\hbox{\cmsss Z\kern-.36em Z}}
 {\lower1.2pt\hbox{\cmsss Z\kern-.36em Z}}\else{\cmss Z\kern-.4em
 Z}\fi}
\def\IZ{\relax\ifmmode\mathchoice
{\hbox{\cmss Z\kern-.4em Z}}{\hbox{\cmss Z\kern-.4em Z}}
{\lower.9pt\hbox{\cmsss Z\kern-.4em Z}}
{\lower1.2pt\hbox{\cmsss Z\kern-.4em Z}}\else{\cmss Z\kern-.4em
Z}\fi}

\def\narrowplus{\kern -.04truein + \kern -.03truein}
\def\narrowminus{- \kern -.04truein}
\def\narrowminussub{\kern -.02truein - \kern -.01truein}

\def\kh{K\"{a}hler}

\def\a{{\alpha}}
\def\g{{\gamma}}

\def\l{{\lambda}}

\def\r{{\rightarrow}}

\def\frac#1#2{{#1\over #2}}

\def\com#1#2{{ \left[ #1, #2 \right] }}

\def\acom#1#2{{ \left\{ #1, #2 \right\} }}

\def\IZ{\relax\ifmmode\mathchoice
{\hbox{\cmss Z\kern-.4em Z}}{\hbox{\cmss Z\kern-.4em Z}}
{\lower.9pt\hbox{\cmsss Z\kern-.4em Z}}
{\lower1.2pt\hbox{\cmsss Z\kern-.4em Z}}\else{\cmss Z\kern-.4em
Z}\fi}
\def\IB{\relax{\rm I\kern-.18em B}}
\def\IC{{\relax\hbox{$\inbar\kern-.3em{\rm C}$}}}
\def\ID{\relax{\rm I\kern-.18em D}}
\def\IE{\relax{\rm I\kern-.18em E}}
\def\IF{\relax{\rm I\kern-.18em F}}
\def\IG{\relax\hbox{$\inbar\kern-.3em{\rm G}$}}
\def\IGa{\relax\hbox{${\rm I}\kern-.18em\Gamma$}}
\def\IH{\relax{\rm I\kern-.18em H}}
\def\II{\relax{\rm I\kern-.18em I}}
\def\IK{\relax{\rm I\kern-.18em K}}
\def\IP{\relax{\rm I\kern-.18em P}}

\font\cmss=cmss10 \font\cmsss=cmss10 at 7pt
\def\IR{\relax{\rm I\kern-.18em R}}

\def\f{\psi}
\def\l{\lambda}

\def\s{{\sigma}}
\def\ss{{\tilde s}}
\def\1{{\bf 1}}
\def\3{{\bf 3}}
\def\4{{\bf 4}}
\def\5{{\bf 5}}
\def\7{{\bf 7}}
\def\2{{\bf 2}}
\def\8{{\bf 8}}

%

%
%
\def\eqnn#1{\xdef #1{(\secsym\the\meqno)}\writedef{#1\leftbracket#1}%
\global\advance\meqno by1\wrlabeL#1}
\def\eqna#1{\xdef #1##1{\hbox{$(\secsym\the\meqno##1)$}}
\writedef{#1\numbersign1\leftbracket#1{\numbersign1}}%
\global\advance\meqno by1\wrlabeL{#1$\{\}$}}
\def\eqn#1#2{\xdef #1{(\secsym\the\meqno)}\writedef{#1\leftbracket#1}%
\global\advance\meqno by1$$#2\eqno#1\eqlabeL#1$$}



\lref\rsegal{G. Segal and A. Selby, \cmp{177}{1996}{775}. }
\lref\rdorold{N. Dorey, V. Khoze, M. Mattis, D. Tong and S. Vandoren, 
hep-th/9703228, \np{502}{1997}{59}.}
\lref\rsen{A. Sen, hep-th/9402002, Int. J. Mod. Phys. {\bf A9} (1994) 3707;
hep-th/9402032, \pl{329}{1994}{217}. }
\lref\rcallias{C. Callias, \cmp{62}{1978}{213}\semi E. Weinberg, 
\prd{20}{1979}{936}.}
\lref\rblum{J. Blum, hep-th/9401133, \pl{333}{1994}{92}.}
\lref\rsgreen{M. B. Green and S. Sethi, hep-th/9808061.}
\lref\rbateman{H. Bateman, ed. A. Erdelyi, 
{\it Higher Transcendental Functions, vol. 2}, McGraw-Hill Book Company, 1953. }
\lref\ryi{P. Yi, hep-th/9704098, \np{505}{1997}{307}.}
\lref\rsavmark{S. Sethi and M. Stern, hep-th/9705046, \cmp{194}{1998}{675}.}
\lref\rwitthmon{E. Witten, hep-th/9511030, \np{460}{1996}{541}.}
\lref\rgg{M. B. Green and M. Gutperle, hep-th/9711107, JHEP 01 (1998) 05.}
\lref\rgreg{G. Moore, N. Nekrasov and S. Shatashvili, hep-th/9803265.}
\lref\rSS{S. Sethi and L. Susskind, hep-th/9702101,
    \pl{400}{1997}{265}.}
\lref\rBS{T. Banks and N. Seiberg,
    hep-th/9702187, \np{497}{1997}{41}.}
\lref\rreview{N. Seiberg, hep-th/9705117.}
\lref\rpp{J. Polchinski and P. Pouliot, hep-th/9704029, \prd{56}{1997}{6601}.}
\lref\rds{M. Dine and N. Seiberg, hep-th/9705057, \pl{409}{1997}{239}.}
\lref\rdorey{N. Dorey, V. Khoze and M. Mattis, hep-th/9704197, 
\np{502}{1997}{94}.}
\lref\rbf{T. Banks, W. Fischler, N. Seiberg and L. Susskind, hep-th/9705190, 
\pl{408}{1997}{111}.}
\lref\rlowe{D. Lowe, hep-th/9810075.}

\lref\rK{N. Ishibashi, H. Kawai, Y. Kitazawa and A. Tsuchiya, hep-th/9612115.}
\lref\rCallias{C. Callias, Commun. Math. Phys. {\bf 62} (1978), 213.}
\lref\rPD{J. Polchinski, hep-th/9510017, \prl{\bf 75}{1995}{47}.}
\lref\rWDB{E. Witten,  hep-th/9510135, Nucl. Phys. {\bf B460} (1996) 335.}
\lref\rSSZ{S. Sethi, M. Stern, and E. Zaslow, Nucl. Phys. {\bf B457} (1995)
484.}
\lref\rGH{J. Gauntlett and J. Harvey, Nucl. Phys. {\bf B463} 287. }
\lref\rAS{A. Sen, Phys. Rev. {\bf D53} (1996) 2874; Phys. Rev. {\bf D54} (1996)
2964.}
\lref\rWI{E. Witten, Nucl. Phys. {\bf B202} (1982) 253.}
\lref\rPKT{P. K. Townsend, Phys. Lett. {\bf B350} (1995) 184.}
\lref\rWSD{E. Witten, Nucl. Phys. {\bf B443} (1995) 85.}
\lref\rASS{A. Strominger, Nucl. Phys. {\bf B451} (1995) 96.}
\lref\rBSV{M. Bershadsky, V. Sadov, and C. Vafa, Nucl. Phys. {\bf B463}
(1996) 420.}
\lref\rBSS{L. Brink, J. H. Schwarz and J. Scherk, Nucl. Phys. {\bf B121}
(1977) 77.}
\lref\rCH{M. Claudson and M. Halpern, Nucl. Phys. {\bf B250} (1985) 689.}
\lref\rSM{B. Simon, Ann. Phys. {\bf 146} (1983), 209.}
\lref\rGJ{J. Glimm and A. Jaffe, {\sl Quantum Physics, A Functional Integral
Point of View},
Springer-Verlag (New York), 1981.}
\lref\rADD{ U. H. Danielsson, G. Ferretti, B. Sundborg, Int. J. Mod. Phys. {\bf
A11} (1996) 5463\semi   D. Kabat and P. Pouliot, Phys. Rev. Lett. {\bf 77}
(1996), 1004.}
\lref\rDKPS{ M. R. Douglas, D. Kabat, P. Pouliot and S. Shenker,
hep-th/9608024,
Nucl. Phys. {\bf B485} (1997), 85.}
\lref\rhmon{S. Sethi and M. Stern, hep-th/9607145, Phys. Lett. {\bf B398} 
(1997), 47.}
\lref\rBFSS{T. Banks, W. Fischler, S. H. Shenker, and L. Susskind, 
hep-th/9610043,
Phys. Rev. {\bf D55} (1997) 5112.}
\lref\rBHN{ B. de Wit, J. Hoppe and H. Nicolai, Nucl. Phys. {\bf B305}
(1988), 545\semi
B. de Wit, M. M. Luscher, and H. Nicolai, Nucl. Phys. {\bf B320} (1989),
135\semi
B. de Wit, V. Marquard, and H. Nicolai, Comm. Math. Phys. {\bf 128} (1990),
39.}
\lref\rT{ P. Townsend, Phys. Lett. {\bf B373} (1996) 68.}
\lref\rLS{L. Susskind, hep-th/9704080.}
\lref\rFH{J. Frohlich and J. Hoppe, hep-th/9701119.}
\lref\rAg{S. Agmon, {\it Lectures on Exponential Decay of Solutions of
Second-Order Elliptic Equations}, Princeton University Press (Princeton) 1982.}
\lref\rY{P. Yi, hep-th/9704098.}
\lref\rDLhet{ D. Lowe, hep-th/9704041.}
\lref\rqm{R. Flume, Ann. Phys. {\bf 164} (1985) 189\semi
M. Baake, P. Reinecke and V. Rittenberg, J. Math. Phys. {\bf 26} (1985) 1070.}
\lref\rbb{K. Becker and M. Becker, hep-th/9705091, \np{506}{1997}{48}\semi
K. Becker, M. Becker, J. Polchinski and A. Tseytlin, hep-th/9706072,
\prd{56}{1997}{3174}.}
\lref\rpw{J. Plefka and A. Waldron, hep-th/9710104, \np{512}{1998}{460}.}
\lref\rhs{M. Halpern and C. Schwartz, hep-th/9712133, Int. J. Mod. Phys. {\bf A13} (1998)
4367.}

\lref\rlimit{N. Seiberg hep-th/9710009, \prl{79}{1997}{3577}\semi
A. Sen, hep-th/9709220.}
\lref\rentin{D.-E. Diaconescu and R. Entin, hep-th/9706059,
\prd{56}{1997}{8045}.}
\lref\rgreen{M. B. Green and M. Gutperle, hep-th/9701093, \np{498}{1997}{195}.}
\lref\rpioline{B. Pioline, hep-th/9804023.}
\lref\rgl{O. Ganor and L. Motl, hep-th/9803108.}
\lref\rds{M. Dine and N. Seiberg, hep-th/9705057, \pl{409}{1997}{209}.}
\lref\rberg{E. Bergshoeff, M. Rakowski and E. Sezgin, \pl{185}{1987}{371}.}
\lref\rBHP{M. Barrio, R. Helling and G. Polhemus, hep-th/9801189.}
\lref\rper{P. Berglund and D. Minic, hep-th/9708063, \pl{415}{1997}{122}.}
\lref\rspin{P. Kraus, hep-th/9709199, \pl{419}{1998}{73}\semi
J. Harvey, hep-th/9706039\semi
J. Morales, C. Scrucca and M. Serone, hep-th/9709063, \pl{417}{1998}{233}.}
\lref\rdine{M. Dine, R. Echols and J. Gray, hep-th/9805007; hep-th/9810021.}
\lref\rber{D. Berenstein and R. Corrado, hep-th/9702108, \pl{406}{1997}{37}.}
\lref\rnonpert{A. Sen, hep-th/9402032, \pl{329}{1994}{217}; hep-th/9402002,  
Int. J. Mod. Phys. {\bf
A9} (1994) 3707\semi
N. Seiberg and E. Witten, hep-th/9408099, \np{431}{1995}{484}.
}
\lref\rberkooz{M. Berkooz and M. R. Douglas, hep-th/9610236, 
\pl{395}{1997}{196}.}
\lref\rmike{M. R. Douglas, hep-th/9612126, J.H.E.P. 9707:004, 1997.}
\lref\rhitchin{N. Hitchin, math.DG/9909002.}
\lref\rjost{J. Jost and K. Zuo, ``Vanishing Theorems for $L^2$-cohomology on
infinite coverings of compact \kh\ manifolds and applications in algebraic
geometry,'' preprint no. 70, Max-Planck-Institut f\"ur Mathematik in den Naturwissenschaften,
Leipzig (1998).}

\lref\rpss{S. Paban, S. Sethi and M. Stern, hep-th/9805018, \np{534}{1998}{137}.}
\lref\rpsst{S. Paban, S. Sethi and M. Stern, hep-th/9806028. }
\lref\rperiwal{V. Periwal and R. von Unge, hep-th/9801121.}
\lref\rfer{M. Fabbrichesi, G. Ferreti and R. Iengo, hep-th/9806018.}
\lref\rkac{V. G. Kac and A. Smilga, hep-th/9908096.}
\lref\rporrati{M. Porrati and A. Rozenberg, hep-th/9708119, \np{515}{1998}{184}.}
\lref\rgg{M. B. Green and M. Gutperle, hep-th/9804123, \prd{58}{1998}{46007}.}
\lref\rmoore{G. Moore, N. Nekrasov and S. Shatashvili, hep-th/9803265, 
\cmp{209}{2000}{77}.}
\lref\rkrauth{W. Krauth, H. Nicolai and M. Staudacher, hep-th/9803117, \pl{431}{1998}{31}; 
W. Krauth and M. Staudacher, hep-th/9804199, \pl{435}{1998}{350}.}
\lref\rkonech{A. Konechny, hep-th/9805046, J.H.E.P. 9810:018, 1998.}
\lref\rvanhove{P. Vanhove, hep-th/9903050, Class. Quant. Grav. {\bf 16} (1999) 3147.}
\lref\rfrolich{J. Frohlich, G. M. Graf, D. Hasler, J. Hoppe and S.-T. Yau, 
hep-th/9904182; G. M. Graf and J. Hoppe, hep-th/9805080. }
\lref\rqm{M. Claudson and M. Halpern, \np{250}{1985}{689}\semi
R. Flume, Ann. Phys. {\bf 164} (1985) 189\semi
M. Baake, P. Reinecke and V. Rittenberg, J. Math. Phys. {\bf 26} (1985) 1070.}

\Title{\vbox{\hbox{hep-th/0001189}
\hbox{DUK-CGTP-00-03, IASSNS--HEP--00/118}}}
{\vbox{\centerline{Invariance Theorems for Supersymmetric}
\vskip8pt\centerline{Yang-Mills Theories}}}

\centerline{Savdeep
Sethi$^\ast$\footnote{$^1$} {sethi@sns.ias.edu} and Mark
Stern$^\dagger$\footnote{$^2$} {stern@math.duke.edu} }

\medskip\centerline{$\ast$ \it School of Natural Sciences, Institute for
Advanced Study, Princeton, NJ 08540, USA}
\medskip\centerline{$\dagger$ \it Department of Mathematics, Duke University,  
Durham, NC 27706, USA}

\vskip 0.5in

We consider quantum mechanical Yang-Mills theories with eight supercharges and 
a $Spin(5) \times SU(2)_R $ flavor symmetry. We 
show that all normalizable ground states in these gauge theories
are invariant under this flavor symmetry. This includes, as a special case, all 
bound states of D0-branes and D4-branes. As a consequence, all bound states of
D0-branes are invariant under the $Spin(9)$ flavor symmetry. When combined with
index results, this implies that the bound state of two D0-branes is unique.

\vskip 0.1in
\Date{1/00}

\newsec{Introduction}

The existence of normalizable vacua in supersymetric 
Yang-Mills theories is a question that arises in many different contexts in string
theory and field theory. Index arguments can be used to determine whether any
vacua exist, but not exactly how many vacua. An index only counts the difference
between the number of bosonic and fermionic vacua. To count the actual number of 
vacua, we need more information such as how the vacua transform under the global 
symmetries of the theory. 

In this paper, we consider quantum mechanical Yang-Mills theories with eight
supercharges and an $Spin(5) \times SU(2)_R $ symmetry. 
We take our theories to be dimensional 
reductions of $d=6$ N=1 Yang-Mills theories coupled 
to matter. The question of normalizable ground states in these models arises in the
study of bound states of D0-branes and D4-branes \refs{\rberkooz, \rDKPS}; 
for example, a single D0-brane and 
a single D4-brane can be shown to bind using $L^2$ index arguments \rhmon\ 
generalized to theories without a gap.  
Other examples from string theory involve D0-branes moving on orbifolds \rmike, and 
the question of counting H-monopoles in the heterotic string 
\refs{\rwitthmon, \rhmon}.

In the following section, we describe the field content and symmetries of these
gauge theories. We then show that all normalizable ground states in these 
theories must be invariant under the $SU(2)_R$ symmetry. The argument we give is
suggested by recent work on the $L^2$-cohomology of hyper\kh\
spaces by Hitchin \rhitchin. Our result should have implications for defining and computing 
the $L^2$-cohomology 
of instanton moduli spaces. Certain instanton moduli spaces appear as 
Higgs branches in gauge theories of the kind under consideration. For example, 
the moduli space of $U(N)$ instantons in $\IR^4$ appears as the Higgs branch of 
the quantum mechanics describing D0-D4 systems. 
Although these spaces can be singular, their embedding into quantum mechanical gauge 
theory provides a natural regularization of the singularities. Heuristically, 
the wavefunction for
a state corresponding to a form on the Higgs branch is smoothed out by leaking onto
the Coulomb branch. It would be interesting to explore this connection further. 

There is a second $R$-symmetry in these theories which comes from the dimensional
reduction of the Lorentz group. For reductions of $d=6$ N=1 Yang-Mills theories, 
this is 
a $Spin(5)$ symmetry.  Using basically the same argument as in the case of the $SU(2)_R$
symmetry, we show that all normalizable ground states in these theories
are invariant under this $Spin(5)$ symmetry. 
For reductions of $d=10$ N=1 Yang-Mills theories \rqm, the $R$-symmetry group is
$Spin(9)$. It is quite straightforward to argue that as a consequence
of the $SU(2)_R \times Spin(5)$ invariance theorem, all ground states
in these theories with sixteen supercharges  must be invariant under the
$Spin(9)$ symmetry. 

We can couple these invariance theorems with results from $L^2$ index theory 
\refs{\rsavmark, \ryi}. The 
$L^2$ index for the non-Fredholm theory\foot{By non-Fredholm, we mean a theory without
a gap.} of two D0-branes is proven to be one \rsavmark. 
We also know that the $L^2$ index for the theory
of a single D0-brane and a single D4-brane is one \rhmon. Our invariance results
imply that all bound states in these theories are bosonic, and therefore unique. 
These results can also be combined with other interesting but heuristic attempts 
to study the $L^2$ index by either deforming the Yang-Mills theory  
\refs{\rporrati, \rkac}, or by using insights from string theory \rgg\ 
to compute the bulk and defect terms. The bulk terms for various Yang-Mills theories
have been directly computed in \refs{\rmoore, \rvanhove,\rkrauth}. 
There have also 
been a number of comments on the implications of invariance for the asymptotic form
of particular bound state wavefunctions \refs{\rhs, \rfrolich}.

\newsec{The Field Content and Symmetries}
\subsec{The vector multiplet supercharge}

The argument we wish to make requires reasonably little explicit knowledge of
the gauge theory. There is a $Spin(5) \times SU(2)_R$ symmetry which 
commutes with the Hamiltonian $H$. 
Since we are considering a gauge theory, we must have at least one vector
multiplet. It contains five scalars $x^\mu$ with $\mu=1,\ldots, 5$ transforming 
in the 
$({\bf 5}, {\bf 1})$ of the symmetry group. These scalars transform in the adjoint
representation of the gauge group $G$. Let $p^\mu$ be the associated 
canonical momenta obeying,
\eqn\bosquant{ \com{x^\mu_A}{p^\nu_B} = i\delta^{\mu\nu} \delta_{AB}, }
where the subscript $A$ is a group index. 

Associated to these bosons are eight real fermions $\l_a$ where $a=1,\ldots,8$  
transforming in the $( \4, \2)$ representation of the symmetry group. These fermions
are also in the adjoint representation of the gauge group. The eight
supercharges also transform in the $( \4, \2)$ representation.
These fermions obey the usual quantization relation, 
\eqn\fermquant{ \acom{\l_{aA}}{\l_{bB}} = \delta_{ab} \delta_{AB}.} 
Let $\g^\mu$ be hermitian real 
gamma matrices which obey, 
\eqn\gmat{ \acom{\g^\mu}{\g^\nu} = 2 \delta^{\mu\nu}. }
Appendix A includes an explicit basis for these gamma matrices along with a 
discussion of the symmetry group action. 

The supercharge takes the form,
\eqn\super{ Q_{a}^v = \left( \g^\mu p^\mu_A \l_A \right)_a +{1\over 2} f_{ABC} 
\left( \g^{\mu\nu} \l_A x^\mu_B x^\nu_C \right)_a
+ D_{ab A} \l_{b A},}
where $f_{ABC}$ are the structure constants and $ \g^{\mu \nu} = 
(1/2)( \g^{\mu} \g^{\nu} - \g^{\nu} \g^{\mu}).$
The real anti-symmetric matrix $D$ does not involve momenta. The $D$-term
transforms in the $(\1,\3)$ representation of the symmetry group, and in the adjoint
representation of the gauge group. The precise form of $D$ is not 
important for our argument. In general, there can be
many vector multiplets. In that case, the terms in the supercharge \super\ 
generalize in an obvious way. 

\subsec{The hypermultiplet supercharge}

A hypermultiplet contains four real scalars which we can package into a 
quaternion
$q$ with components $q^i$ where $i=1,2,3,4$. This field transforms 
as $({\bf 1}, {\bf 2})$ under the symmetry group, and in some representation $T$ 
of the gauge group. We again introduce canonical 
momenta
$p_i$ satisfying the usual commutation relations. Now $SU(2)_R \sim Sp(1)_R$ 
is the group of unit quaternions. We choose $SU(2)_R$ to act on a hypermultiplet
$q$ by right multiplication by a unit quaternion. The gauge symmetry commutes with
the $SU(2)_R$ symmetry and acts by left multiplication on $q$. See Appendix A for
a more detailed discussion.

The superpartner to $q$ is a real fermion $\f_a$ with $a=1,\ldots, 8$ 
satisfying,
\eqn\secondquant{ \acom{\f_{a}^R}{\f_{bS}} = \delta_{a b} \delta^R_{S}.}
These fermions transform in the $({\bf 4}, {\bf 1})$ representation, and the $R,S$
subscripts index the $T$ representation of $G$. For $n$ hypermultiplets, the gauge group
$G$ acts via a subgroup of the $Sp(n)_L$ symmetry. In terms of the 
$s^j$
operators given in Appendix A, the hypermultiplet charge takes the form
\eqn\freehyper{  Q^{h}_a = s^j_{ab} \f_{b} \, p_{j} + I_{ab} \f_b.}
We have lumped all the interactions into the non-derivative operator $I$ which
transforms in the $\2$ of $SU(2)_R$. We also need to note that $I$ is proportional to 
$x^\mu \g^\mu$ with a proportionality constant that commutes with the $Spin(5)$ generators. 
We have also suppressed gauge indices. 
Note that since the $s^j$ implement right multiplication by a quaternion, they 
commute with $\g^\mu$. Again, there can be many hypermultiplets in different representations
of the gauge group. In that case, the 
hypermultiplet
supercharge \freehyper\ generalizes in a straightforward way. 
The full Hermitian supercharge is the sum of the vector and hypermultiplet supercharges, 
$$ Q_a =  Q_{a}^v + Q^{h}_a. $$

\subsec{The $SU(2)_R$ currents}

The three generators of $SU(2)_R$ correspond to right multiplication by $I,J,K$
and are given in terms of the gauge invariant rotation generators,
\eqn\defW{ W_{ij} =  q_i p_j - q_j p_i.}
Again here and in the subsequent discussion, we generally suppress
gauge indices. In accord with prior notation, we denote the three $SU(2)_R$
 generators by $\ss^i$:
\eqn\defss{\eqalign{ \ss^2 & = W_{12} - W_{34} + {i\over 2}\, \l s^2 \l \cr
\ss^3 & = W_{13} + W_{24} + {i\over 2}\, \l s^3 \l \cr
\ss^4 & = W_{14} - W_{23} + {i\over 2}  \, \l s^4 \l. }}
As they should, these generators act on the bosons of the hypermultiplet and the
fermions of the vector multiplet. Adding either more vector multiplets or more
hypermultiplets is straightforward: we simply need to sum the contributions to
the three currents \defss\ from each multiplet. 

\subsec{The $Spin(5)$ currents}

The ten generators of $Spin(5)$ act on the bosons of the vector multiplet and all
fermions in the problem. The generators are given by:
\eqn\genfive{ T^{\mu\nu} = x^\mu p^\nu - x^\nu p^\mu - {i\over 4} \g^{\mu \nu}_{ab}
\left( \l_a \l_b + \f_a \f_b \right).}
Adding either more vector multiplets or more
hypermultiplets is again straightforward.

\newsec{An Invariance Argument for the $SU(2)_R$ Symmetry}

\subsec{Relating the $SU(2)_R$ currents to the supercharge}

A key point in the argument is a relation between the supercharge and
the $SU(2)_R$ currents. For some choice of $v^i_a$, we want to show that:
\eqn\relation{ \ss^i = \sum_a \, \acom{Q_a}{v^i_a}.}
Let us start with the vector multiplet. We take a candidate gauge singlet, 
\eqn\vvector{ (v_1)^i_a = \left( s^i \g^\nu \l \right)_a x^\nu. }
First note that this choice anti-commutes
with $Q^h$ because $\l$ anti-commutes with $\f$.  It also anti-commutes with the
$D$-term in \super. To see this, we compute:
\eqn\quickcheck{ \eqalign{ \sum_a \acom{D_{ab} \l_b }{(v_1)^i_a} & = x_A^\nu \tr \left(
s^i \g^\nu D_A^T \right), \cr}}
However, we can immediately see that \quickcheck\ vanishes by noting that 
the operator $s^i \g^\nu D^T$ does not contain a singlet under $Spin(5)$.
The trace of the operator therefore vanishes. Our choice for $v_1$ anti-commutes with 
$ {1\over 2} f_{ABC} 
\left( \g^{\mu\nu} \l_A x^\mu_B x^\nu_C \right)_a$ for the same reason: 
the resulting trace does not contain a singlet of $Spin(5)$.

What remains is the following anti-commutator which is not hard to compute,
\eqn\vectsurvive{ \eqalign{ \sum_a \acom{\left( \g^\mu p^\mu \l \right)_a}{(v_1)^i_a} & 
\sim  i \, \l s^i \l.}}
The exact proportionality constant does not matter for this argument. The important
point is that we can use \vvector\ to generate the terms in the $SU(2)_R$ currents
which act on vector multiplets. 

For the hypermultiplet, we take the following candidate gauge singlet:
\eqn\vhyper{ (v_2)^i_a = \left( s^i s^l \f \right)_a q^l.}
Note that $v_2$ anti-commutes with $Q^v$ because $\l$ anti-commutes with $\f$. 
It is also not too hard to argue that the anti-commutator of $v_2$ with the 
interaction term $I$ in \freehyper\ must vanish. We see that, 
\eqn\scheck{ \eqalign{ \sum_a \acom{I_{ab} \f_b }{(v_2)^i_a} & \sim
 q^l \tr \left( s^i s^l I \right), \cr}}
but $ s^i s^l I$ does not contain a singlet under the $Spin(5)$ action on fermions
 because $I$ is proportional
to $ \gamma^\mu$ so the trace vanishes. 

Again what remains is the anti-commutator, 
\eqn\hypersurvive{ \sum_a \acom{s^j_{ab} \f_{b} \, p_{j}}{(v_2)^i_a}.}
It is easy to check that the $\f \f$ terms in the anti-commutator vanish because, 
$$ \sum_k \, \f \{ s^k \}^T s^i s^k \f =0.$$ 
With a little
additional work, we find that \hypersurvive\ gives precisely the bosonic terms in
\defss\ up to an overall non-vanishing constant. We therefore conclude that for appropriately
chosen constants $\a_1$ and $\a_2$, the choice
\eqn\pickv{ v^i_a = \a_1 (v_1)^i_a  + \a_2 (v_2)^i_a }
satisfies \relation. 

\subsec{Rotating a ground state}

We assume there exists a normalizable ground state $\Psi$ which is not a singlet
under $SU(2)_R$. Under some $SU(2)_R$ rotation, we obtain another non-trivial $L^2$
zero-energy state. What does $L^2$ imply? Let us collectively denote all the bosonic
coordinates $x$ and $q$ by $y^i$ where $i=1, \ldots, D$. Normalizability requires 
that,
$$ < \Psi, \Psi> =  \int{d^D y \, \Psi^\dagger (y^i) \, \Psi(y^i) } < \infty. $$
For some $\ss^i$, the state $ \ss^i \Psi$ is a non-trivial ground state. 
It satisfies the relation,  
\eqn\obvious{ Q_a \left( \ss^i \Psi \right) = Q_a \Psi = 0, }
for each $a$ by definition of a ground state.  Using \relation, we find that
\eqn\manips{ \eqalign{ \ss^i \Psi & =   \sum_a \, \acom{Q_a}{v^i_a} \Psi, \cr
& = \sum_a Q_a \left( v^i_a \Psi \right). }} 
The new ground state looks $Q$-trivial. To show that it really is physically trivial, 
we need to check that it has zero norm. Since $Q$ is Hermitian and kills $\ss^i \Psi$,
the norm of $\ss^i \Psi$  vanishes  if we can integrate by 
parts. To integrate by parts, we argue as in Jost and Zuo \refs{
\rjost, \rhitchin}: in terms of $y = |y^i|$, we can cutoff of the integral using a 
smooth bump function $ \rho_R (y)$
which vanishes for $y>2R$, satisfies $|d\rho_R| < 4/R$ and is one for $y<R$,
$$ <\ss^i \Psi,  \ss^i \Psi>= \lim_{R \r \infty}< \rho_R(y) \ss^i \Psi,  \ss^i \Psi>.$$
Using \obvious\ and \manips, we see that
\eqn\mmanips{ \eqalign{
 <\ss^i \Psi,  \ss^i \Psi> & = \lim_{R \r \infty} < \rho_R(y) \ss^i \Psi, \sum_a \, 
\acom{Q_a}{v^i_a} \Psi >, \cr
& =  \lim_{R \r \infty} \sum_a < \com{Q_a}{\rho_R(y)} \,\ss^i \Psi, v^i_a \Psi >. \cr}}
We see that $\com{Q_a}{\rho_R(y)}$ is $O(1/y)$ and vanishes for
$y<R$ and  $y> 2R$. Since $ v_i^a$ is $O(y)$ at worst, 
the right hand side of \mmanips\ vanishes. The $SU(2)_R$ symmetry therefore
acts trivially on all normalizable ground states.

\newsec{Invariance Under the $Spin(5)$ Symmetry}
\subsec{Relating the $Spin(5)$ currents to the supercharge}

We want to use essentially the same argument as in the $SU(2)_R$ case. 
 For some choice of $v^{\mu\nu}_a$, we want to show that:
\eqn\frelation{ T^{\mu\nu} = \sum_a \, \acom{Q_a}{v^{\mu\nu}_a}.}
Let us start with the vector multiplet. We take a candidate gauge singlet, 
\eqn\fvvector{ (v_1)^{\mu\nu}_a = \left\{ \g^\mu x^\nu - \g^\nu x^\mu \right\}_{ab} \l_b. }
Again this choice anti-commutes
with $Q^h$ because $\l$ anti-commutes with $\f$. The anti-commutator with
 ${1\over 2} f_{ABC} 
\left( \g^{\mu\nu} \l_A x^\mu_B x^\nu_C \right)_a$ results in a trace of three gamma
matrices and so vanishes. It also anti-commutes with the
$D$-term in \super. To see this, we compute:
\eqn\fquickcheck{ \eqalign{ \sum_a \acom{D_{ab} \l_b }{(v_1)^{\mu\nu}_a} & = D_{ab} 
\left\{ \g^\mu x^\nu - \g^\nu x^\mu \right\}_{ab}. \cr}}
However, this combination does not contain a singlet under $Spin(5)$ so  
\fquickcheck\ vanishes. 

We are left with the following anti-commutator which we need to compute quite carefully,
\eqn\fvectsurvive{ \eqalign{ \sum_a \acom{\left( \g^\mu p^\mu \l \right)_a }{
(v_1)^{\mu\nu}_a} & 
=  8 \left( x^\nu p^\mu - x^\mu p^\nu \right) + 2 i \l \g^{\mu\nu} \l.}}
This computation is sensitive to the size of the $\g$ matrix. We obtain 
precisely the right ratio between the bosonic and fermion terms in \fvectsurvive\ 
because the theory is reduced from six dimensions. We would not obtain the right
ratio had we considered a theory reduced from ten dimensions with a $Spin(9)$ symmetry.
Again, we can use \fvvector\ to generate the terms in the $Spin(5)$ currents
which act on vector multiplets. 

For the hypermultiplet, we take the following choice:
\eqn\fvhyper{ (v_2)^{\mu\nu}_a = \left( \g^{\mu\nu} s^i \f \right)_a q^i.}
Again $v_2$ anti-commutes with $Q^v$ because $\l$ anti-commutes with $\f$. 
In much the same way as before, we can argue that the anti-commutator of $v_2$ with the 
interaction term $I$ in \freehyper\ must vanish. We see that, 
\eqn\fscheck{ \eqalign{ \sum_a \acom{I_{ab} \f_b }{(v_2)^{\mu\nu}_a} & \sim
 q^i \tr \left( I \g^{\mu\nu} s^i \right), \cr}}
but $I \g^{\mu\nu} s^i $ again does not contain a singlet under $Spin(5)$ 
so the trace vanishes. 

The remaining anti-commutator involves the kinetic term in the hypermultiplet charge, 
\eqn\fhypersurvive{ \sum_a \acom{s^j_{ab} \f_{b} \, p_{j}}{(v_2)^{\mu\nu}_a} = - i \f 
\g^{\mu\nu} \f.}
Again we conclude that for appropriately
chosen constants $\a_1$ and $\a_2$, the choice
\eqn\fpickv{ v^{\mu\nu}_a = \a_1 (v_1)^{\mu\nu}_a  + \a_2 (v_2)^{\mu\nu}_a }
satisfies \frelation. 
A straightforward repeat of the argument given in section {\it 3.2} then implies that the
$Spin(5)$ symmetry acts trivially on all normalizable ground states.

\subsec{Theories with sixteen supercharges}

For theories obtained by reduction from ten dimensions, 
the previous 
argument does not apply directly to the $Spin(9)$ symmetry for reasons mentioned earlier.
These theories contain scalars $y^i$ where $i=1,\ldots,9$
transforming in the adjoint representation of the gauge group. The superpartners to 
these scalars are real fermions $\eta_\a$ where $\a=1,\ldots,16$ also in the
adjoint representation. 

However, we can always view these theories as special cases of theories with eight
supercharges. We choose any $5$ of the $9$ scalars $y^i$ to be the vector multiplet,
and the remaining $4$ scalars comprise an adjoint hypermultiplet. Of the original
$Spin(9)$ symmetry, only a $Spin(5)\times SU(2)_L \times SU(2)_R$ subgroup is manifest. 
The scalars
decompose in the following way, 
\eqn\decompsc{ {\bf 9}  \quad \r \quad ( \5 , \1, \1) \oplus (\1, \2, \2).}
The fermions decompose according to, 
\eqn\decompfm{ {\bf 16} \quad \r \quad ( \4, \1, \2) \oplus (\4, \2, \1).}
Our invariance argument implies that all normalizable ground states are invariant
under the  $Spin(5)\times SU(2)_R$ symmetry. However, this is true regardless of how
we embed  $Spin(5)\times SU(2)_R$ into $Spin(9)$. This is only possible if the full
$Spin(9)$ symmetry acts trivially on all normalizable ground
states.

\bigbreak\bigskip\bigskip\centerline{{\bf Acknowledgements}}\nobreak
The work of S.S. is supported  by the William Keck Foundation and by 
NSF grant PHY--9513835; that of M.S. by NSF grant DMS--9870161.

\vfill\eject

\appendix{A}{Quaternions and Symplectic Groups}

We will summarize some useful relations between quaternions and symplectic 
groups.
Let us label a basis for our quaternions by $\{ \1 , I, J, K\}$ where,
$$I^2=J^2=K^2=-\1, \qquad IJK = - \1. $$
A quaternion $q$ can then be expanded in components
$$ q = q^1 + I q^2 + J q^3 + K q^4. $$
The conjugate quaternion $\bar{q}$ has an expansion
$$ q = q^1 - I q^2 - J q^3 - K q^4. $$
The symmetry group $Sp(1)_R \sim SU(2)_R$ is the group of unit quaternions. Let 
$\Lambda $ be a field
transforming in the $\2$ of $Sp(1)_R$. If we view $Sp(1)_R$ acting on $\Lambda$ 
as
right multiplication by a unit quaternion $g$ then,
$$ \Lambda \,\r\, \Lambda g. $$
In this formalism, $\Lambda$ is valued in the quaternions. 
Equivalently, we can expand $\Lambda$ in components and express the action of 
$g$ in the following way,
$$ \Lambda_a \,\r\, g_{ab} \Lambda_b, $$
where $g_{ab}$ implements right multiplication by the unit quaternion $g$.
For example, right multiplication by $I$ on $q$ gives
$$ \eqalign{ q &\,\r\, q I \cr
               & \, \r \,  q^1 I - q^2 - q^3 K + q^4 J,}$$
which can be realized by the matrix
\eqn\defs{  I^R = \pmatrix{0 & -1 & 0 & 0 \cr 
            1 & 0 & 0 & 0 \cr
            0 & 0 & 0 & 1 \cr
            0 & 0 & -1 & 0 }}
acting on $q$ in the usual way $ q_a \, \r\, I^R_{ab} \, q_b$. 
The matrices $J^R$ and $K^R$ realize right multiplication by $J,K$
while ${\1}^R$ is the identity matrix:
\eqn\defstwo{  J^R = \pmatrix{0 & 0 & -1 & 0 \cr 
            0 & 0 & 0 & -1 \cr
            1 & 0 & 0 & 0 \cr
            0 & 1 & 0 & 0 }, \qquad
             K^R = \pmatrix{0 & 0 & 0 & -1 \cr 
            0 & 0 & 1 & 0 \cr
            0 & -1 & 0 & 0 \cr
            1 & 0 & 0 & 0 }. }
We define operators $s^j$ in terms of $\left\{ {\1}^R, I^R, J^R, K^R \right\}$
$$   s^1 = \pmatrix{\1^R & 0 \cr 0 & \1^R}, \quad
                s^2 = \pmatrix{I^R & 0 \cr 0 & I^R}, \quad
                 s^3 = \pmatrix{J^R & 0 \cr 0 & J^R}, \quad
                s^4 = \pmatrix{K^R & 0 \cr 0 & K^R}.$$

In a similar way, the group $Sp(2) \sim Spin(5)$ is the group of 
quaternion-valued
$2\times 2$ matrices with unit determinant. We will
view $Sp(2)$ as acting by left multiplication on a field $\Psi$ in the defining
representation. So an
element $U\in Sp(2)$ acts in the following way:
$$ \Psi \,\r\, U \Psi.$$
Equivalently, in terms of components
$$ \Psi_a \,\r\, U_{ab} \Psi_b. $$   
Lastly, we can give an explicit form for the gamma matrices \gmat\ in terms of 
quaternions:
$$ \g^1 = \pmatrix{1 & 0 \cr 0 & -1 }, \qquad
\g^2 = \pmatrix{0 & 1 \cr 1 & 0 }, \qquad \g^3 = \pmatrix{0 & I \cr -I & 0 } $$
$$ \g^4 = \pmatrix{0 & J \cr -J & 0 }, \qquad \g^5 = \pmatrix{0 & K \cr -K & 0 
}.$$
In turn, $\{ I, J, K \} $ can be expressed in terms of the Pauli matrices $\s^i$
$$ \s^1 = \pmatrix{0 & 1 \cr 1 & 0}, \qquad 
\s^2 = \pmatrix{0 & -i \cr i & 0}, \qquad \s^3 = \pmatrix{1 & 0 \cr 0 & -1} $$ 
as $4\times 4$ real anti-symmetric matrices: 
$$ I = \pmatrix{0 & \s^1 \cr -\s^1 & 0 }, \qquad
 J = \pmatrix{-i\s^2 & 0 \cr 0 & -i \s^2 }, 
\qquad K = \pmatrix{0 & \s^3 \cr -\s^3 & 0 }. $$

\vfill\eject	

\listrefs
\bye